# Emergence of new magic numbers N = 16 and 34 through tensor interaction in Skyrme Hartree Fock theory

Rupayan Bhattacharya*
University of Calcutta
92 A.P.C. Road, Kolkata 700009, India

Melting of N=20 shell and development of N = 16 and N = 34 shells for neutron-rich nuclei have been studied extensively through inclusion of tensor interaction in Skyrme Hartree Fock theory optimized to reproduce the splitting $\Delta 1f$ of $^{40,\,48}$Ca and $^{56}$Ni nuclei. Evolution of gap generated by the energy difference of single particle levels $\nu 2s_{1/2}$ and $\nu 1d_{3/2}$ has been found to be responsible for the development of a shell closure at N = 16. The splitting pattern of spin-orbit partners 2p shell model state in Ca, Ti, Cr, Fe and Ni isotopes indicates formation of a new shell at N = 34 region.

---

*rup_bhat@hotmail.com,, Tel: 91 9830273693

## I. INTRODUCTION

The existence of magic numbers characterizes the shell structure of atomic nuclei. Reproduction of magic numbers [1] after introduction of a spin-orbit interaction was a great achievement from the point view of theoretical nuclear physics. All the properties related to shell closure are relevant to beta-stable nuclei found in the Segre chart. But as one approaches dripline, there are dramatic changes in shifting and regrouping of single particle levels producing new values of magic numbers. New areas of magicity appear while conventional shell closures melt away. The study of exotic nuclei with completely different N/Z ratios from the nuclei of established beta stable region poses an interesting challenge to nuclear structure physics. The experimental signatures of magic numbers are indicated by relatively high energy of the first excited state and a quite small B(E2,$2^+ \to 0^+_{gs}$) value in some even-even nuclei. There is another parameter which provides a pointer to magic nuclei, viz., "shell gap" which is the energy difference between the last hole state below the Fermi surface and first particle state above the Fermi surface of the nucleus under consideration. Change in the number of protons (-neutrons) in the atomic nucleus changes the energy of the single particle states, and, consequently, the size of the shell gaps also alters due to the interaction between protons and neutrons lying in different orbitals. Otsuka *et al.* [2-4] have pointed out the possible role of the monopole component of the nucleon-nucleon interaction in determining the magnitude of the change. A dramatic shift of the single particle energies is predicted as one moves from $^{30}$Si to $^{24}$O due to the strong tensor interaction between the $d_{5/2}$ protons and the $d_{3/2}$ neutrons [2]. Disappearance of N=20 shell and development of N=16 shell for neutron-rich nuclei have been studied extensively from different experimental aspects [5-7]. Abrupt ending of the chain of oxygen isotopes at A = 24 (N = 16) and existence of a longer chain of Flourine isotopes up to A = 31 is believed to be the result of a large shell gap generated by the energy difference of single particle levels ν$2s_{1/2}$ and ν$1d_{3/2}$. Recently Tshoo et al [5] have investigated the unbound excited states of neutron-rich $^{24}$O via proton inelastic scattering in inverse kinematics. The excitation energy of the first excited state was determined to be 4.65 ± 0.14 MeV, and the spin parity was assigned $J^\pi = 2^+$. In addition, a relatively small β$_2$ parameter was determined (0.15 ± 0.04), indicative of the spherical character of $^{24}$O and the large shell gap at N = 16. The low-lying level structure of $^{24}$O was studied via proton knockout from $^{26}$F by Hoffman et al. [6]. They reported a higher excitation energy (E = 4.72 MeV) for the first excited state of $^{24}$O than that of $^{22}$O (E = 3.20 MeV) [7]. Elekes et al [8] have measured neutron single particle energies in $^{23}$O using the $^{22}$O(d, p)$^{23}$O → $^{22}$O + *n* process. The energies of the resonant states have been deduced to be 4.00(2) MeV and 5.30(4) MeV. The measured 4.0 MeV energy difference between the $2s_{1/2}$ and $1d_{3/2}$ states gives the size of the *N* =16 shell gap which is in agreement with the recent USD05 (''universal'' sd from 2005) shell model calculation, and is large enough to explain the unbound nature of the oxygen isotopes heavier than A =24. Fernandez-Dominguez et al have shown from the spectroscopic study of 21O that the large energy difference obtained between the 3/2+ and 1/2+ states indicates the emergence of the N = 16 shell gap estimated to be 5.1 (11) MeV.

This oxygen anomaly could not be reproduced by shell model calculations. Otsuka et al [2-4] have shown that inclusion of 3N force instead of NN interaction produces the neutron dripline at $^{24}$O correctly. Based on surveys of the neutron separation energies $S_n$ and the interaction cross sections σ$_I$ for the neutron-rich p-sd and the sd shell region Ozawa et al [9] have shown the creation of a new magic number at N = 16. From results obtained with a phenomenological one-boson exchange potential (OBEP) two-body interaction Brown and Richter [10] have shown the emergence of shell closure at neutron number 16. They have also emphasized the failure of reproduction of neutron number dependence of single particle energies by conventional H-F models.

Evolution of shell structures of the neutron-rich Ca and Ni nuclei is an interesting topic. Though the conventional magic numbers 28 and 50 remain valid for stable nuclei, the situation changes dramatically for neutron rich unstable nuclei. A new magic number at $N = 32$ has been indicated by the experiments on $^{52}$Ca [11-12]. Beta-decay studies of $^{56}$Cr indicated a possible sub-shell gap at neutron number N=32 [12]. Evidence for a subshell closure at $N=32$ was seen first in the Ca isotopes, where both the mass [13] and relatively high value of the excitation energy of the first excited $2^+$ [$E(2_1^+)$] in $^{52}$Ca [14] suggested added stability. However, systematic data for isotopes beyond $^{52}$Ca are unavailable, and the appearance of a subshell gap at $N=32$ was reinforced by the proton shell closure at $Z=20$ as shown by Tondeur [15]. The value $E(2_1^+)$ =1007 keV of $^{56}$Cr$_{32}$ is more than 100 keV above that observed in neighbors $^{54,58}$Cr.

The systematic behaviour of $E(4_1^+)/E(2_1^+)$ in the Cr isotopes is also suggestive of a subshell closure at $N=32$ [16]. Liddick et al [17] have shown a subshell closure at N = 32 from the systematic variation of $E(2_1^+)$ along the $N=32$ isotonic chain completed for the $\pi f_{7/2}$ nuclides with the determination of $E(2_1^+)$ =1495 keV for $^{54}$Ti [18].

From the intermediate energy Coulomb excitation on $^{52-56}$Cr isotopes and absolute B(E2,0+ → $2_1^+$) transition rates measurement Dinca et al [19] confirmed the presence of a subshell closure at neutron number N = 32 in neutron rich nuclei above the doubly closed shell nucleus $^{48}$Ca. Their data did not show any shell closure at N = 34.

On the theoretical side shell-model calculations using the effective interaction GXPF1 by Honna et al [20] revealed the possibility of sub-shell gaps at neutron numbers N=32 and N=34. Later Nakada [21] investigated shell structure of the neutron-rich Ca and Ni nuclei by applying the self-consistent Hartree-Fock (HF) calculations with the semi-realistic NN interactions which indicated shell closure at N = 32.

In view of the above evidences it appears that a comprehensive study of formation of shells in neutron rich nuclei around N = 20 and N = 28 is required. In this paper we shall show that inclusion of tensor interaction within the Skyrme-Hartree-Fock approach and optimization of the tensor coupling constants for a good reproduction of spin-orbit splitting of ν1f shell model state of $^{40,48}$Ca and $^{56}$Ni can give us clear signatures of new areas of magicity for exotic nuclei rich with neutrons. Perturbative treatment to include tensor interaction has been avoided as per the suggestion of Bender et al [22] who have discussed about the instabilities involved in the method. The physics of the present approach has been tested exhaustively in the shell evolution across the nuclear chart [23-24].

The organization of the paper is as follows: Mathematical formalism for the work will be developed in section 2. Results will be discussed in section 3. Finally a brief summary of the work and conclusion will be presented in section 4.

## II. MATHEMATICAL FORMALISM

In Hartree-Fock theory the tensor interaction is given by

$$V_T = \frac{T}{2} \left\{ \left[(\boldsymbol{\sigma}_1 \cdot \boldsymbol{k}')(\boldsymbol{\sigma}_2 \cdot \boldsymbol{k}') - \frac{1}{3}(\boldsymbol{\sigma}_1 \cdot \boldsymbol{\sigma}_2)k'^2 \right] \delta(\boldsymbol{r}_1 - \boldsymbol{r}_2) + \delta(\boldsymbol{r}_1 - \boldsymbol{r}_2)\left[(\boldsymbol{\sigma}_1 \cdot \boldsymbol{k})(\boldsymbol{\sigma}_2 \cdot \boldsymbol{k}) - \frac{1}{3}(\boldsymbol{\sigma}_1 \cdot \boldsymbol{\sigma}_2)k^2 \right] \right\}$$

$$+ U\{(\boldsymbol{\sigma}_1 \cdot \boldsymbol{k}')\delta(\boldsymbol{r}_1 - \boldsymbol{r}_2)(\boldsymbol{\sigma}_2 \cdot \boldsymbol{k}) - \frac{1}{3}(\boldsymbol{\sigma}_1 \cdot \boldsymbol{\sigma}_2) \times [\boldsymbol{k}' \cdot \delta(\boldsymbol{r}_1 - \boldsymbol{r}_2)\boldsymbol{k}]\} \tag{1}$$

In order to get the effect of the tensor interaction, the time-even tensor and spin-orbit parts of the energy density function (EDF) is considered,

$$H_T = C_0^J \mathbf{J}_0^2 + C_1^J \mathbf{J}_1^2 \tag{2}$$

$$H_{SO} = C_0^{\nabla J} \rho_0 \nabla \cdot \mathbf{J}_0 + C_1^{\nabla J} \rho_1 \nabla \cdot \mathbf{J}_1 , \tag{3}$$

where

$$\mathbb{J}^2 = \sum_{\rho\sigma} J_{\rho\sigma}^2 , \text{ and} \tag{4}$$

$$J_{\rho\sigma} = \tfrac{1}{3} J^{(0)} \delta_{\rho\sigma} + \tfrac{1}{2} \varepsilon_{\rho\sigma\tau} J_\tau + J_{\rho\sigma}^{(2)} \tag{5}$$

If spherical symmetry is imposed, the scalar part $J^{(0)}$ and symmetric-tensor part $J_{\rho\sigma}^{(2)}$ of spin-current density vanish and one is left with

$$H_T = \tfrac{1}{2} C_0^J J_0^2(r) + \tfrac{1}{2} C_1^J J_1^2(r) \tag{6}$$

$$H_{SO} = -C_0^{\nabla J} J_0(r) \frac{d\rho_0}{dr} - C_1^{\nabla J} J_1(r) \frac{d\rho_1}{dr} \tag{7}$$

Here $C_0^J$, $C_1^J$ and $C_1^{\nabla J}$ are tensor coupling constants. The form of the spin-orbit potential is given by

$$V_{SO}^q = \frac{1}{2r} \left\{ (C_0^J - C_1^J)J_0(r) + 2C_1^J J_q(r) - (C_0^{\nabla J} - C_1^{\nabla J})\frac{d\rho_0}{dr} - 2C_1^{\nabla J}\frac{d\rho_q}{dr} \right\} \mathbf{L} \cdot \mathbf{S} \tag{8}$$

Lesinski et al [25] have shown that the structure of the tensor part of the EDF can be recast as

$$H_T = \tfrac{1}{2}\alpha(\mathbf{J}_n^2 + \mathbf{J}_p^2) + \beta \mathbf{J}_n \cdot \mathbf{J}_p \qquad (9)$$

where

$$\alpha = C_0^J + C_1^J, \quad \beta = C_0^J - C_1^J \qquad (10)$$

In terms of these new coefficients the spin-orbit potential component is given by

$$V_{s.o.}^q = \frac{W_0}{2r}\left(2\frac{d\rho_q}{dr} + \frac{d\rho_{q'}}{dr}\right) + \left(\alpha\frac{J_q}{r} + \beta\frac{J_{q'}}{r}\right) \qquad (11)$$

where $J_{q(q')}(r)$ is the proton or neutron spin-orbit density defined as

$$J_{q(q')}(r) = \frac{1}{4\pi r^3}\sum_i v_i^2(2j+1)\left[j_i(j_i+1) - l_i(l_i+1) - \tfrac{3}{4}\right]R_i^2(r) \qquad (12)$$

In this expression q stands for neutrons(protons) and q' for protons(neutrons), and i = n,l,j runs over all states having the given q(q') and $R_i(r)$ is the radial part of the wave function. $v_i^2$ is the BCS occupation probability of each orbital and

$$C_0^{\nabla J} = -\tfrac{3}{4}W_0 \qquad (13)$$

The spin-orbit current $J_q$ shows shell effect giving strong fluctuations and it gives essentially no contribution for the spin-saturated cases where all spin-orbit partners are either completely occupied or empty. But the contribution from $J_q$ increases linearly with the number of particles if only one of the spin-orbit partners $J_>$ or $J_<$ is filled.

In eq.(11) $\alpha = \alpha_c + \alpha_T$ and $\beta = \beta_c + \beta_T$. In terms of the usual Skyrme force parameters one can express

$$\alpha_c = \tfrac{1}{8}(t_1 - t_2) - \tfrac{1}{8}(t_1 x_1 + t_2 x_2) \qquad (14)$$

$$\beta_c = -\tfrac{1}{8}(t_1 x_1 + t_2 x_2) \qquad (15)$$

Expressions for the tensor contributions are

$$\alpha_T = \tfrac{5}{12}U, \quad \beta_T = \tfrac{5}{24}(T+U), \qquad (16)$$

where the coupling constants T and U denote the strength of the triplet-even and triplet-odd tensor interactions respectively.. According to Stancu et al [26] the optimal parameters of tensor interaction, viz., $\alpha_T$ and $\beta_T$ should be found from a triangle in the two dimensional ($\alpha_T$, $\beta_T$) plane, lying in the quadrant of negative $\alpha_T$ and positive $\beta_T$. A schematic pairing force has been used through the energy functional

$$E_{pair} = \sum_q G_q \left[ \sum_{\beta \in q} \sqrt{w_\beta(1 - w_\beta)} \right]^2, \qquad (17)$$

Where the pairing matrix elements $G_q$ are constant within each species $q \in \{\pi, \nu\}$ and $w_\beta$ are the occupation probabilities of the shell model states.

Selection of a good force parameter set which correctly reproduce the location of the centroid energies of neutron 1i and proton 1h shell model intruder states of $^{208}$Pb responsible for shell closure was done first. The centroid energy is given by

$$E_{cent} = \frac{l+1}{2l+1} E_> + \frac{l}{2l+1} E_< \quad , \quad E_> = E_{j=l+\frac{1}{2}}, \quad E_< = E_{j=l-\frac{1}{2}} \qquad (18)$$

We have compared the calculated values with the experimentally observed values as the detailed calculations by Zalewski et al [27] have shown the impact of combined polarization effects e.g. mass polarization, shape polarization and spin polarization on the spin-orbit splitting is indeed very small reflecting a cancellation of polarization effects exerted on the $j = l \pm \frac{1}{2}$ which guided us to calculate the spin-orbit splitting of shell closed nuclei without considering the mass (time-even), shape (time-even) and spin (time-odd) polarization effects. As the splitting of intruder states produces the shell gap, it is important that the set of force parameters reproduces the observed splitting. We have used SKP (28), Sly5 (29) and SKX(30) and KDEOV [31] parameter sets in our calculation. The reasons behind the use of these sets are: i) the $\mathbf{J}^2$ terms from the central force is taken into account which is necessary for inclusion of the tensor force, ii) they reproduce the ground state properties and the single particle structure of $^{208}$Pb, the heaviest stable doubly magic nucleus, both theoretically and experimentally one of the most well studied nuclei, in a way much better than any other parameter set.

At first variational studies of the spin-orbit splitting of $1f_{7/2} - 1f_{5/2}$ levels for isoscalar spin-saturated N = Z nucleus $^{40}$Ca were performed as per prescription of Zalewski et al [27] to get the optimized spin-orbit coupling constant $C_0^{\nabla J}$. The tensor interaction does not have any contribution because of isospin symmetry. A reduction in values of $W_0$ (-4/3 $C_0^{\nabla J}$) is necessary to decrease the separation energy of the spin-orbit doublet $1f_{7/2} - 1f_{5/2}$ and the reduction required to get reasonable agreement with the experimental results for neutron $\Delta 1f$ varied from ~10% for

SKP parameter set to ~35% for SKX set. To get the optimized isoscalar tensor coupling constant $C_0^J$ focus was shifted to $^{56}$Ni which is a spin-unsaturated isoscalar nucleus and the same shell model states are involved. The splitting of spin-orbit partner $1f_{7/2} - 1f_{5/2}$ was utilised to optimize the coupling constant $C_0^J$. Finally, the isovector tensor coupling constant $C_1^J$ was fixed by reproducing the spin-orbit level splittings of 1f state of isospin asymmetric N ≠ Z nucleus $^{48}$Ca. Optimum adjustment of the tensor coupling constants was done after calculating the spin-orbit splitting of several shell model states near the Fermi-surface of the doubly shell closed nucleus $^{208}$Pb. .Zalewski et al have shown that the total polarization effect is very small in the case of single particle states of $^{208}$Pb. Therefore, theoretically calculated bare single particle energies of the nucleus can be compared directly with the experimental values without taking particle-vibration coupling in to account. These three spin-orbit coupling constants have been used to find out the development of shells in different nuclei.

### III. RESULTS AND DISCUSSION

Reproduction of centroid energies of neutron 1i and proton 1h states responsible for shell closure of $^{208}$Pb is a crucial test for inclusion of tensor interaction in the Skyrme Hartree Fock theory. In Table I we present the results for the force parameter sets SKP, SKX and Sly5 when we include tensor interaction. It is our contention that calculation of single particle energies from any theory should be compared with experimental energies of $^{208}$Pb first as there shell model single particle states have been identified unequivocally. In Table II the energies of spin-orbit partners $1d_{5/2} - 1d_{3/2}$ proton states and $1f_{7/2} - 1f_{5/2}$ for $^{48}$Ca have been presented, which play an important role in the shell closure for Z = 20 and N = 28. It is observed that the level splitting for ν1f has been reproduced quite well by all the parameter sets but the splitting for proton 1d state has been underestimated. Our emphasis was on reproduction of splitting of 1f state for both the neutron as well as proton cases. For proton 1f splitting in $^{48}$Ca we have obtained a difference of 4.99 MeV for KDEOV [31], 5.03 MeV for SKP and 4.96 MeV for Sly5 whereas the experimental value is 4.94 MeV [32]. This performance vindicates our conjecture.

Spherical nature of $^{16}$O confirmed by a small value of $β_2$ (0.15) by Tshoo et al [5] is the main reason to apply spherical symmetry in the solutions of HF equations for oxygen isotopes. Spherical symmetry has been assumed for calculations to all the nuclei under consideration. Actually the indication of a shell closure at N = 16 was obtained while observing the evolution of shell gap of Z = 8 isotopes in Ref. [22]. The energy difference between the ν2$s_{1/2}$ and ν1$d_{3/2}$ state is a measure of the shell gap for N = 16 isotones. While the experimental shell gap in $^{24}$O is 4.0 MeV [8], the gap found is 4.09 MeV by KDEOV, 4.54 MeV by SKP and 3.91 MeV by Sly5 parameters. Then we have extended our calculations to check the development of a shell gap at N = 16 for C, Ne, Mg and Si isotopes. The results for change of Δsd using Sly5, SKP and KDEOV parameters of C, O, Ne, Mg and Si isotopes with neutron numbers are displayed in Fig.

1 where a clear signature of a strong shell closure at N = 16 is observed. Shell closures can easily be identified by a drop of the binding energy for nuclei with two neutrons beyond a shell closure, followed by a more flat distribution with increasing numbers of neutrons. In Fig. 2 we present the variation of two neutron separation energy ($S_{2n}$). The experimental data has been taken from Audi and Wapstra [33]. Theoretical estimates have been done by using SKP parameter set after inclusion of tensor interaction. From the graphs of $S_{2n}$ vs. neutron number it can be observed that there is a clear change in the gradient at N = 16 indicating a shell closure. The shell gap can be evaluated in another way [34] : for the neutron gaps at $N_{magic}$ we have calculated the single-particle energies of the last occupied (below) and the first unoccupied (above) orbit, *b* and *a*, as

$e_b(Z, N_{magic}) = − [E(Z, N_{magic}) − E(Z, N_{magic − 1})] = −S_n(Z, N_{magic})$, (19)
$e_a(Z, N_{magic},) = −[E(, N Z_{magic + 1}) − E(Z, N_{magic})] = −S_n(Z, N_{magic + 1})$, (20)

where *E* are the binding energies and $S_n$ are the neutron separation energies. The energy of the gap is then evaluated as

$E_{gap}(Z, N_{magic}) = e_a − e_b.$ (21)

Alternatively we can write $E_{gap}(Z, N_{magic}) = - E(Z, N_{magic+1}) + 2E(Z, N_{magic}) − E(Z, N_{magic-1})$ 
(22)

The results are displayed in Fig. 3 from where it is quite apparent that a shell closure is appearing at N =16 for O, Ne and Mg nuclei in case of both SKP and SKX parameter sets. Thus it can be seen that inclusion of tensor interaction in Skyrme Hartree Fock theory reorganizes the single particle level ordering for neutron rich nuclei in A ~ 20 – 30 region in such a way that a new area of magicity appears at N = 16 for these nuclei.

Indication of a neutron shell closure at N = 32 was obtained in the theoretical study of shell gap evolution of Z = 20 isotopes in Ref. 23. After filling up of $1f_{7/2}$ shell closure excess neutrons fill up $2p_{3/2}$ state first and then the $2p_{1/2}$ state get filled up. Shell gap at N = 32 is developed due to a large splitting of the 2p shell model state in neutron rich nuclei. The strength of spin-orbit potential of the exotic nuclei plays a vital role in forming an area of magicity in this region. The role of tensor interaction in producing a suitable spin-orbit potential needed to produce a considerably large splitting of 2p states has been probed in this work. In Fig. 4 the energy difference of $2p_{3/2} − 2p_{1/2}$ states of Ca, Ti, Cr, Fe and Ni isotopes using Sly5, SKP and SKX parameter sets is presented. It can be seen that a marked change in the level splitting indicating formation of an island of magicity appears at N = 32 for all these nuclei. This phenomenon vindicates the importance of tensor interaction in shell formation.

That the variation of two neutron separation energy with increasing neutron number provides information about shell closure was discussed earlier and the procedure has been applied to Ca, Ti, Cr, Fe and Ni isotopes to check whether any indication can be found for these nuclei when more and more neutrons are added. In Fig. 5 we find that theoretical results closely follow the experimental values [33] in predicting a shell closure at N = 32. It must be mentioned here that the signature of shell closure is not that distinct.

Shell gaps for N = 32 using binding energy prescription have been evaluated for Ca and Ti nuclei by SKP and SKX parameters. The results appearing at Fig. 6a and 6b shows clearly sharp peaks at N = 32 indicating neutron shell closure for these nuclei. Actually tensor included HF + RPA type of calculation is required to check the energy of first $2^+$ state, other excited states and the corresponding BE(2) values for completing the structure information.

## IV. CONCLUSION

Development of new shell structure in exotic nuclei needs prodding from different angles. Formation of shell closures at N = 16 and at N = 32 drew attentions from several quarters. Introduction of tensor interaction in SKHF theory optimized to reproduce observed splitting of 1f shell model state in $^{40, 48}$Ca and $^{56}$Ni has been found to be able in locating experimentally observed new shell closures at the right position for several neutron rich nuclei. Different mechanisms like variation of splitting of shell model states with increasing neutron number, systematics of two neutron separation energies and difference in energies of the last hole state and first particle state calculated through Skyrme Hartree Fock theory after inclusion of optimized tensor interaction point towards the new areas of magicity at N = 16 and N = 32 convincingly.

**ACKNOWLEDGEMENT** The author thanks the University Grants Commission for support by the Emeritus Fellowship [ No. F.6-34/2011(SA – II)].

Table I

Centroid energies of neutron/proton states responsible for shell closure of $^{208}$Pb

| $E_{nl}$/Force | SKPT[a] | SKXT[b] | SlyT5[c] | EXPT.[d] |
|---|---|---|---|---|
| ν 1i | 5.75 | 6.25 | 5.84 | 6.24 |
| π 1h | 6.31 | 6.77 | 6.85 | 6.83 |

a) Ref. 27  c) Ref. 29  d) Ref. 28  e) Ref. 24

Table II

Separation energies of neutron/proton spin-orbit partner states responsible for shell closure of $^{48}$Ca

| $E_{nl}$/Force | SKPT | SKXT | Sly5T | EXPT.[a] |
|---|---|---|---|---|
| ν Δ1f | 8.06 | 8.15 | 8.03 | 8.01 |
| π Δ1d | 3.70 | 3.18 | 3.44 | 5.29 |

a) Ref. 30

**FIGURE CAPTION**

Fig. 1 "(Colours on line)" Change of $\Delta_{sd}$ of C, O, Ne, Mg and Si isotopes with neutron numbers

Fig. 2 "(Colours on line)" Variation of two neutron separation energy ($S_{2n}$) for Ne, Mg and Si isotopes

Fig. 3 "(Colours on line)" Variation of $E_{gap}$ for O, Ne and Mg nuclei

Fig. 4 "(Colours on line)" Energy difference of $2p_{3/2} - 2p_{1/2}$ states of Ca, Ti, Cr, Fe and Ni isotopes

Fig. 5 "(Colours on line)" Variation of two neutron separation energy ($S_{2n}$) for Ca, Ti, Cr, Fe and Ni isotopes

Fig. 6a "(Colours on line)" Evolution of shell gap for Ca and Ti isotopes by SKP parameters

Fig. 6b "(Colours on line)" Evolution of shell gap for Ca and Ti isotopes by SKX parameters

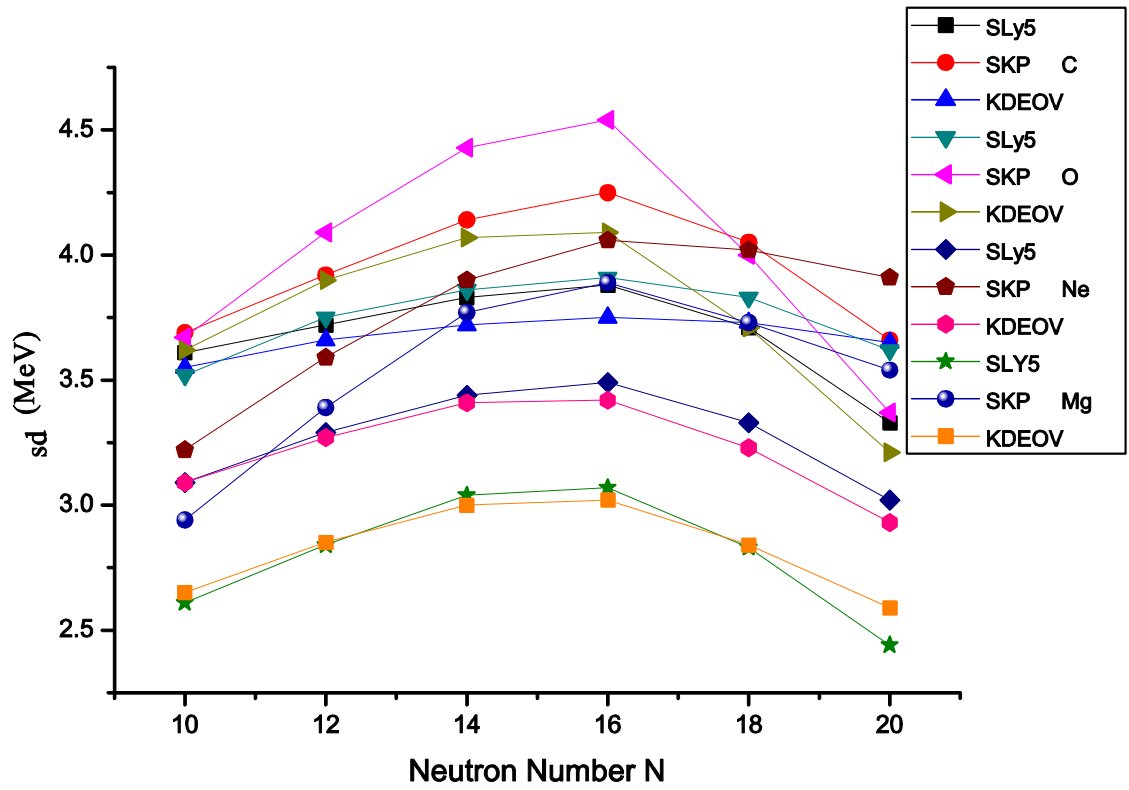

Fig. 1 "(Colours on line)" Change of $\Delta_{sd}$ of C, O, Ne, Mg and Si isotopes with neutron numbers

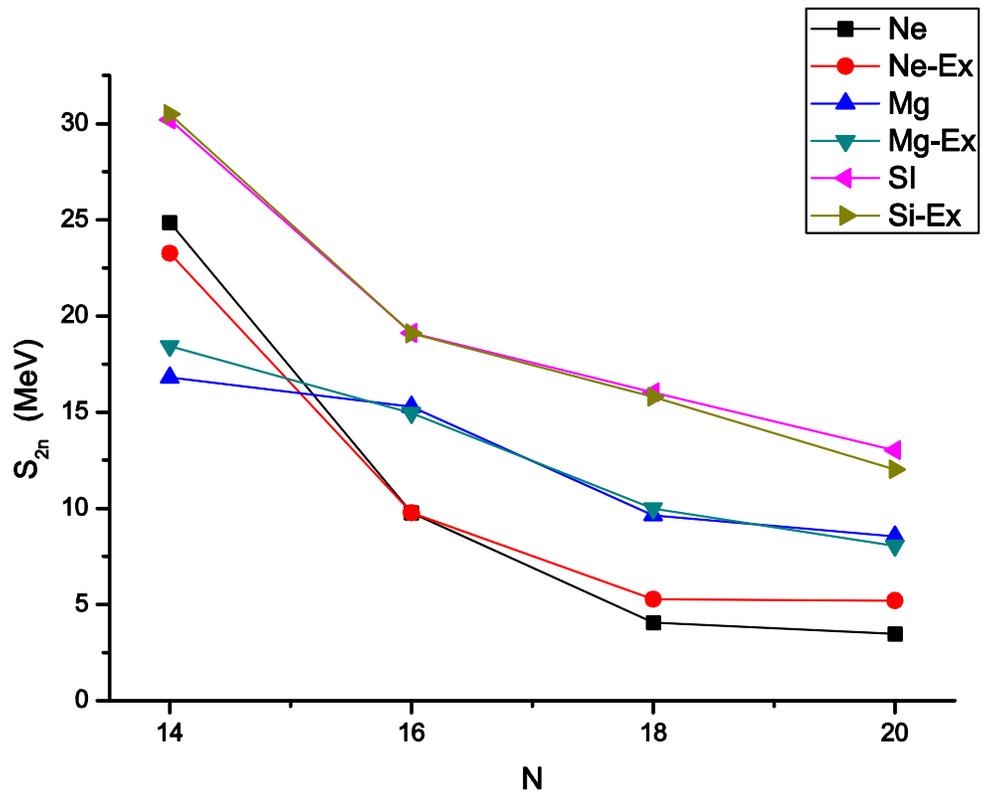

Fig. 2 "(Colours on line)" Variation of two neutron separation energy ($S_{2n}$) for Ne, Mg and Si isotopes

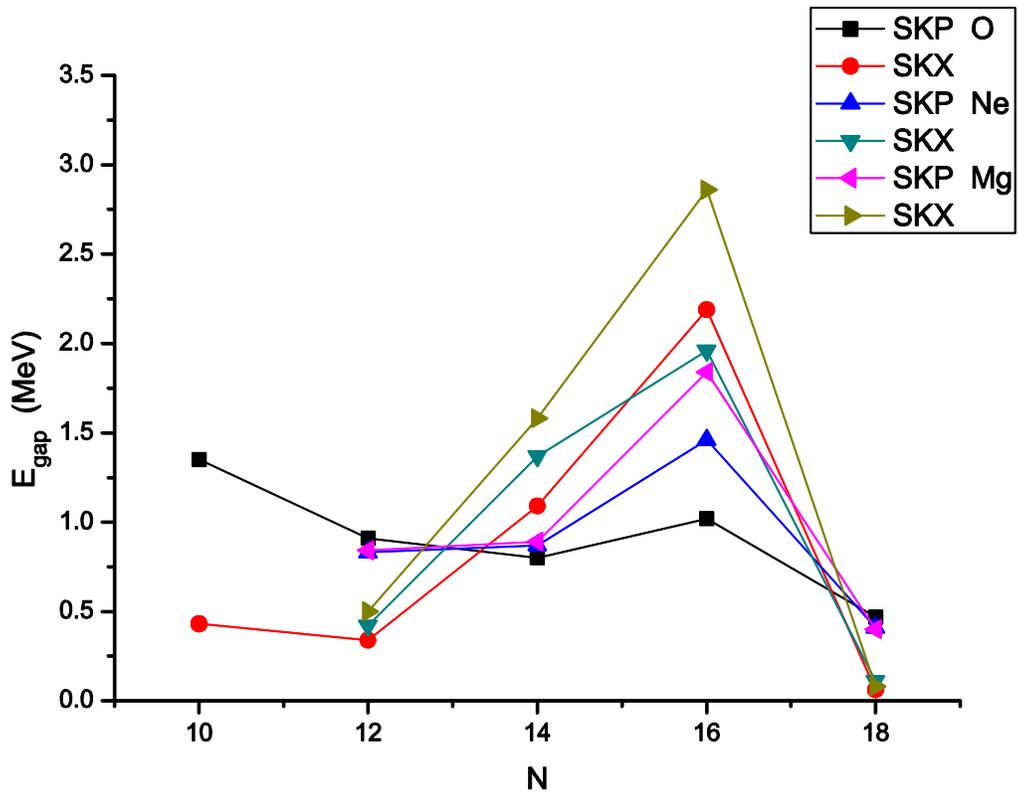

Fig. 3 "(Colours on line)" Variation of $E_{gap}$ for O, Ne and Mg nuclei

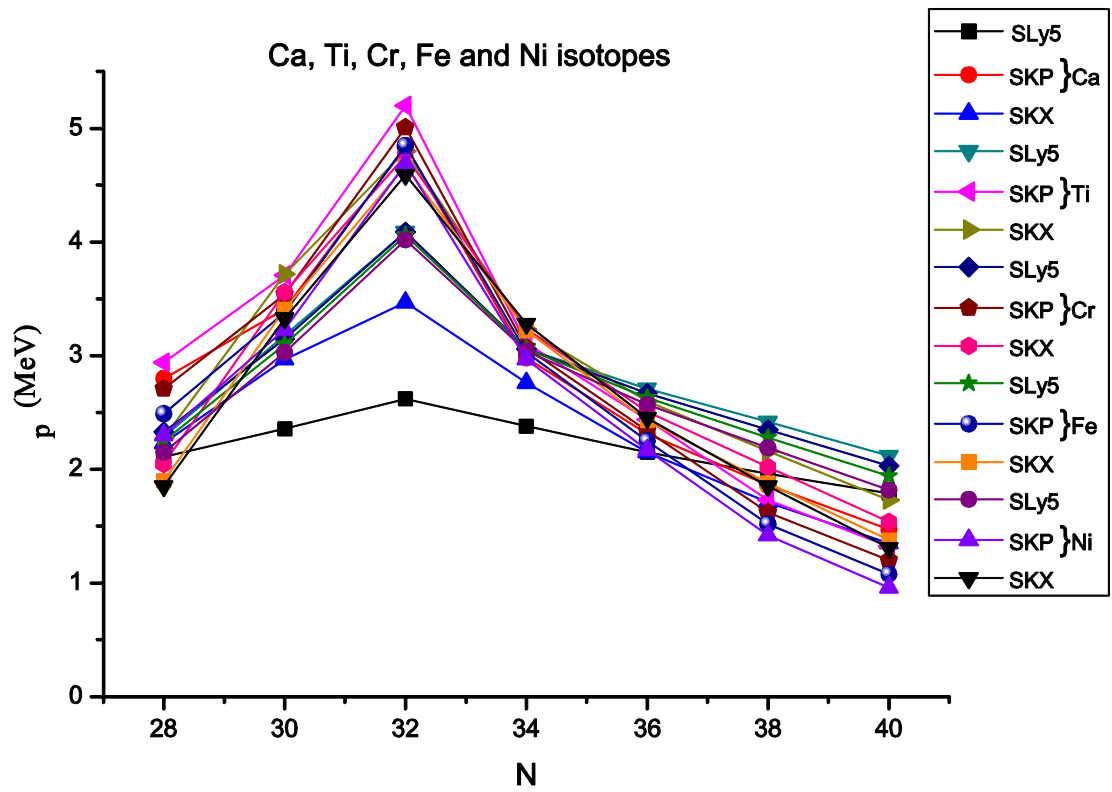

Fig. 4 "(Colours on line)" Energy difference of $2p_{3/2} - 2p_{1/2}$ states of Ca, Ti, Cr, Fe and Ni isotopes

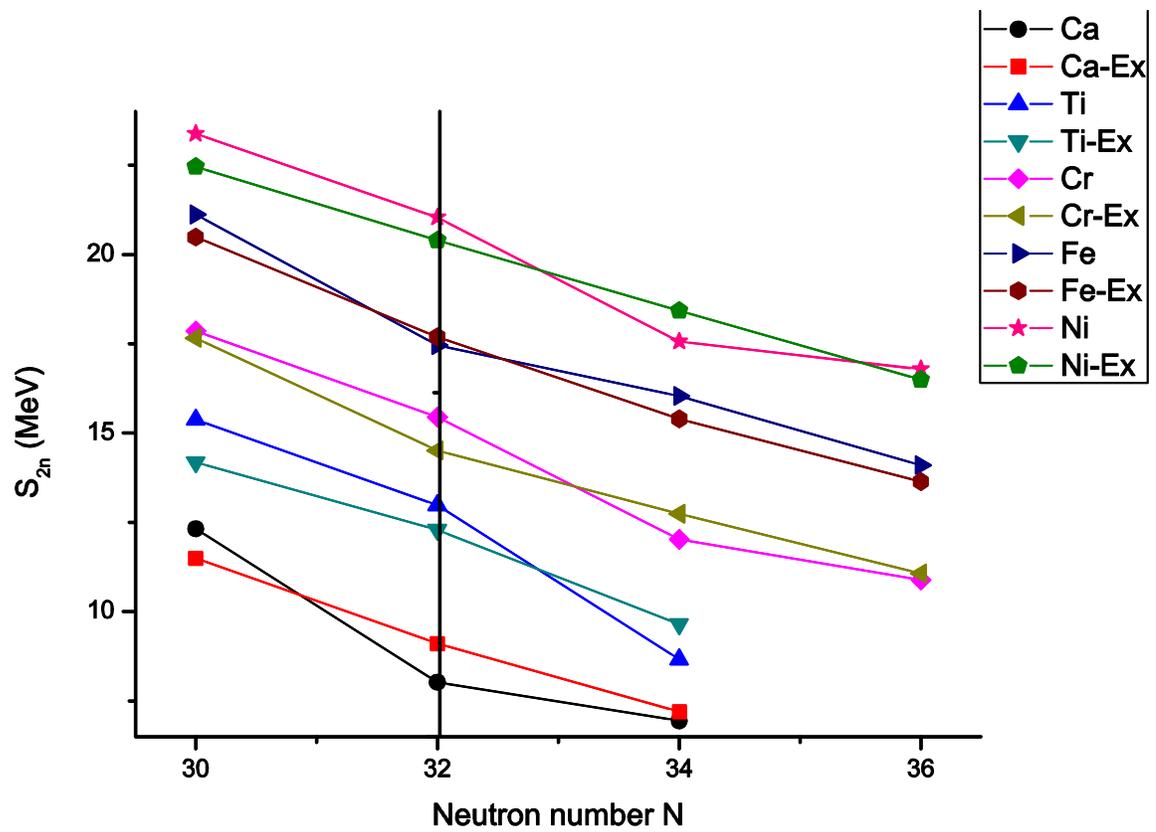

Fig. 5 "(Colours on line)" Variation of two neutron separation energy ($S_{2n}$) for Ca, Ti, Cr, Fe and Ni isotopes

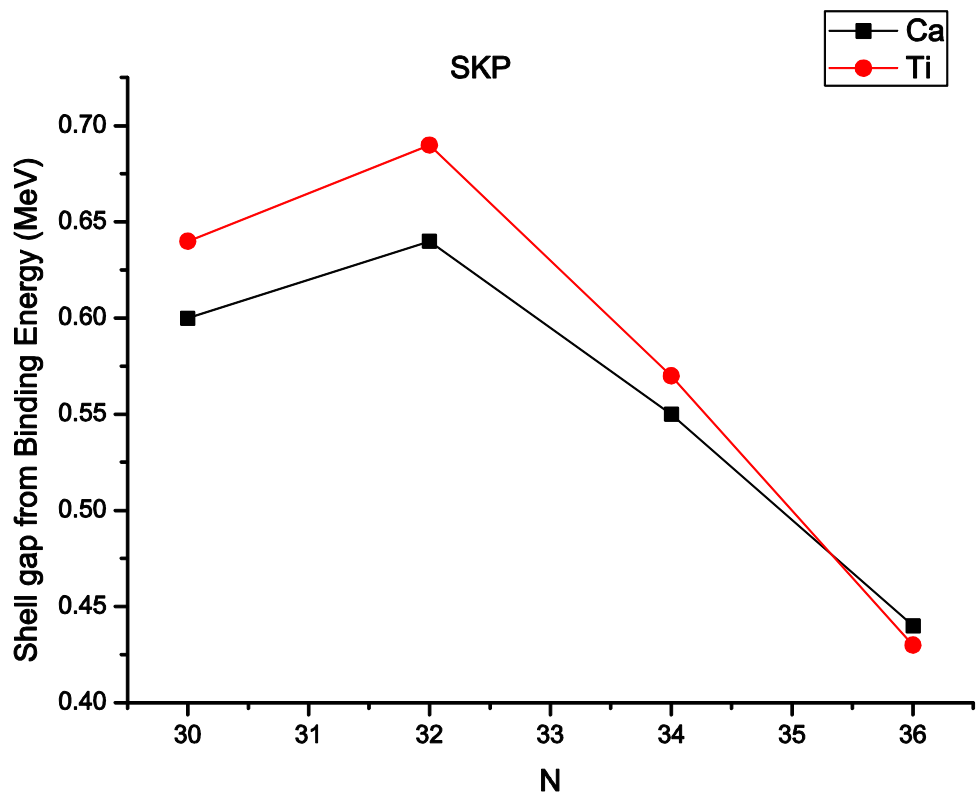

Fig. 6a "(Colours on line)" Evolution of shell gap for Ca and Ti isotopes by SKP parameters

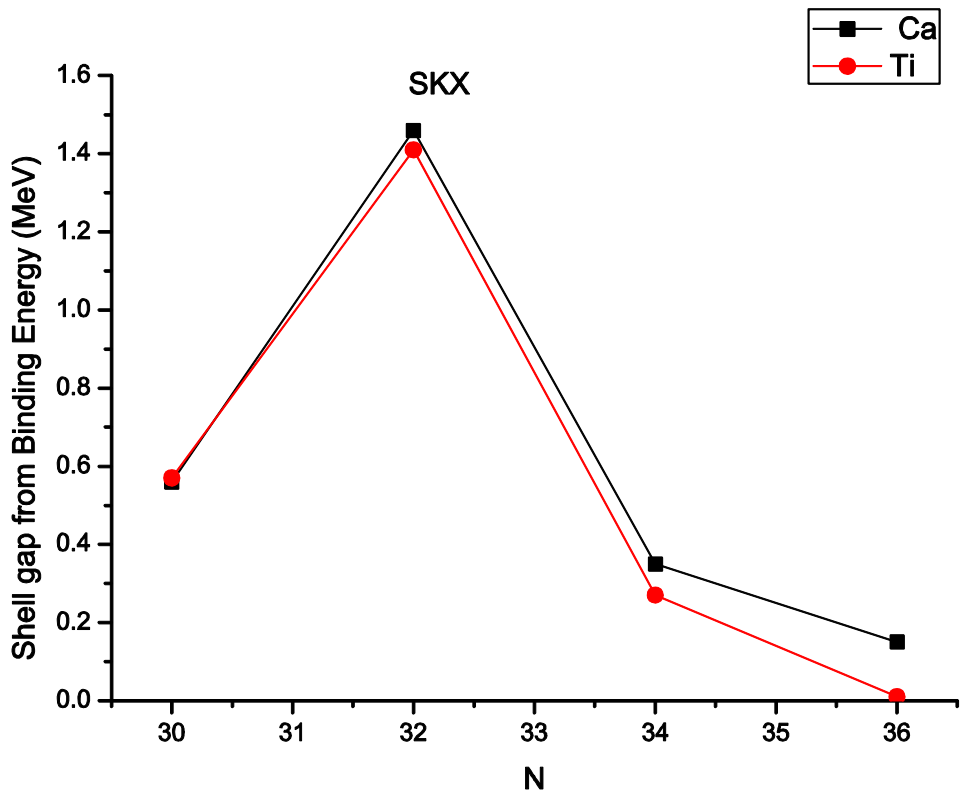

Fig. 6b "(Colours on line)" Evolution of shell gap for Ca and Ti isotopes by SKX parameters